\documentclass[floats,floatfix,amssymb,prd,twocolumn,superscriptaddress,nofootinbib,preprintnumbers]{revtex4-1}

\usepackage{subcaption}
\usepackage{ragged2e}
\DeclareCaptionJustification{justified}{\justifying}
\makeatletter
\newcommand{\subsetsim}{\mathrel{\mathpalette\subset@sim\relax}}
\newcommand{\subset@sim}[2]{%
  \vtop{\offinterlineskip\m@th
    \ialign{\hfil##\cr
      $#1\subset$\cr\noalign{\kern0.5pt}\scalebox{0.9}{$#1\sim$}\cr
    }%
  }%
}
\makeatother

\usepackage{amssymb,amsmath,verbatim,mathtools,needspace,enumitem,etoolbox,graphicx,physics,microtype,afterpage,bm}
\usepackage[dvipsnames, usenames]{xcolor}
\definecolor{linkcolor}{rgb}{0.0,0.3,0.5}
\usepackage{booktabs}
\usepackage[unicode, colorlinks=true, linkcolor=linkcolor, citecolor=linkcolor, filecolor=linkcolor,urlcolor=linkcolor, pdfusetitle]{hyperref}
\usepackage[all]{hypcap}
\usepackage[T1]{fontenc}
\usepackage[utf8]{inputenc}
\usepackage{tabularx}
\usepackage{float}
\usepackage{multirow}
\usepackage{pifont}
\usepackage{lmodern}

\allowdisplaybreaks
\usepackage{tikz}
\usepackage{color}
\usepackage{framed}
\usepackage{hyperref}
\hypersetup{colorlinks, citecolor=bluscuro, linkcolor=black, urlcolor=bluscuro}
\definecolor{rossos}{cmyk}{0,1,1,0.55}
\definecolor{bluscuro}{rgb}{0.15, 0.2, .85}
\definecolor{bluchiaro}{cmyk}{1,.3,0.,0.1}
\definecolor{ForestGreen}{rgb}{0.13, 0.55, 0.13}

\def\bea{\begin{eqnarray}}
\def\eea{\end{eqnarray}}

\def\d{{\mathrm{d}}}
\def\PBH{\text{\tiny{PBH}}}

\newcommand{\bs}{\begin{subequations}}
\newcommand{\es}{\end{subequations}}

\newcommand{\be}{\begin{equation}}
\newcommand{\ee}{\end{equation}}
\renewcommand{\d}{{\rm d}}

\newcommand{\llp}{\left [}
\newcommand{\rrp}{\right ]}
\newcommand{\lp}{\left (}
\newcommand{\rp}{\right )}

\def\lsim{\mathrel{\rlap{\lower4pt\hbox{\hskip0.5pt$\sim$}}
    \raise1pt\hbox{$<$}}}         
\def\gsim{\mathrel{\rlap{\lower4pt\hbox{\hskip0.5pt$\sim$}}
    \raise1pt\hbox{$>$}}}         

\newcommand{\sapienza}{Dipartimento di Fisica, Sapienza Università 
	di Roma, Piazzale Aldo Moro 5, 00185, Roma, Italy}
\newcommand{\infn}{INFN, Sezione di Roma, Piazzale Aldo Moro 2, 00185, Roma, Italy}

\begin{document}

\title{Stochastic gravitational-wave background 
at 3G detectors
\\
as a smoking gun for microscopic dark matter relics}

\author{Gabriele Franciolini}
\email{gabriele.franciolini@uniroma1.it}
\affiliation{\sapienza}
\affiliation{\infn}

\author{Paolo Pani}
\email{paolo.pani@uniroma1.it}
\affiliation{\sapienza}
\affiliation{\infn}


\begin{abstract}
Microscopic horizonless relics could form in the early universe either directly through gravitational collapse or as stable remnants of the Hawking evaporation of primordial black holes. In both cases they completely or partially evade cosmological constraints arising from Hawking evaporation and in certain mass ranges can explain the entirety of the dark matter. We systematically explore the stochastic gravitational-wave background associated with the formation of microscopic dark-matter relics in various scenarios, adopting an agnostic approach and discussing the limitations introduced by existing constraints, possible ways to circumvent the latter, and expected astrophysical foregrounds. Interestingly, this signal is at most marginally detectable with current interferometers but could be detectable by third-generations instruments such as the Einstein Telescope, strengthening their potential as discovery machines.
\end{abstract}

\preprint{ET-0147A-23}

\maketitle

\section{Introduction}
\label{sec:introduction}
The nature of the nonbaryonic dark matter~(DM) that seems to dominate galactic dynamics and accounts for roughly a quarter of the energy content of the universe remains mysterious.
Despite the decade-long experimental effort, DM searches have so far provided only upper bounds on a plethora of models~\cite{BertoneBook,Bertone:2018krk,AlvesBatista:2021gzc}. %
An intriguing possibility is the formation of microscopic compact objects in the early universe, either in the form of primordial horizonless solitons (for example Q-balls~\cite{Coleman:1985ki,Kusenko:1997si}, boson stars~\cite{Jetzer:1991jr,Schunck:2003kk,Liebling:2012fv}, oscillons~\cite{Seidel:1991zh},  fermion soliton stars~\cite{Lee:1986tr,DelGrosso:2023trq}, and fermi-balls \cite{Hong:2020est,Gross:2021qgx}) or primordial black holes~(PBHs)~\cite{Hawking:1974rv,Carr:2020gox} formed during the radiation-dominated era from the collapse of very large inhomogeneities~\cite{Ivanov:1994pa,GarciaBellido:1996qt,Ivanov:1997ia,Blinnikov:2016bxu}. 

According to the standard formation scenario within General Relativity~(GR), PBHs with masses smaller than 
$\sim 10^{-19}\,M_\odot$ (i.e., $\sim10^{14}\,{\rm g}$) 
should be completely evaporated by now and cannot therefore contribute to the DM~\cite{Hawking:1974rv,1975CMaPh..43..199H,Carr:2020gox}. However, this conclusion assumes that GR is valid all the way down to full evaporation, where ultraviolet corrections should become dominant. An intriguing possibility is that ultraviolet effects would eventually halt the evaporation of microscopic PBHs, producing a microscopic relic~\cite{MacGibbon:1987my,Barrow:1992hq,Carr:1994ar,Dalianis:2019asr,Dvali:2020wft} (see~\cite{Chen:2014jwq} for a review).
This can naturally occur in full-blown quantum gravity~\cite{Ashtekar:2005cj,Nicolini:2005vd,Chen:2002tu,Chamseddine:2016ktu} (see~\cite{Hossenfelder:2012jw} for a review), but also in more tractable high-curvature extensions to GR, where BHs feature a minimum mass~\cite{Corelli:2022pio,Corelli:2022phw}.
Likewise, primordial horizonless solitons --~formed within and beyond the Standard Model (e.g.,~\cite{Endo:2022uhe})~-- can evade all the constraints arising from Hawking evaporation~\cite{Carr:2020gox}
and could comprise the entirety of the DM also in mass ranges excluded for PBHs~\cite{Raidal:2018eoo}.
Hawking evaporation is also quenched for nearly-extremal PBHs
(see Ref.~\cite{deFreitasPacheco:2020wdg,deFreitasPacheco:2023hpb} and references therein) and in the presence of large extra dimensions~\cite{Emparan:2000rs,Friedlander:2022ttk}.

Motivated by the above scenarios, in this work we explore the possibility that microscopic horizonless relics could form in the early universe either from the gravitational collapse of large perturbations or as (meta)stable remnants of the Hawking evaporation of PBHs. As we shall argue, in some of the above scenarios these objects are compelling candidates for the entirety of the DM and their formation would be associated with a stochastic gravitational-wave background~(SGWB) potentially observable by future, third-generation~(3G), ground-based interferometers~\cite{Kalogera:2021bya} such as the Einstein Telescope~(ET)~\cite{Hild:2010id,Punturo:2010zz,Maggiore:2019uih} and Cosmic Explorer~\cite{Evans:2016mbw,Essick:2017wyl}. 
The reach of 3G detectors to constrain
the amplitude of a SGWB nonlinearly induced by large scalar perturbations was already mentioned in Ref.~\cite{Romero-Rodriguez:2021aws} (see also Refs.~\cite{Kapadia:2020pir,Kapadia:2020pnr}).
Here we extend the existing discussion connecting the SGWB signal to different scenarios where Hawking remnants and exotic compact objects~\cite{Giudice:2016zpa,Cardoso:2019rvt} may be related to the DM and specifically show that 3G detectors could test the formation of DM relics that partially or totally circumvent the Hawking evaporation bounds.

We shall mostly consider the standard  formation scenario in which primordial compact objects form from the collapse of large inhomogeneities during a radiation dominated era of the universe~\cite{Ivanov:1994pa,GarciaBellido:1996qt,Ivanov:1997ia,Blinnikov:2016bxu}.
This could naturally lead to objects with masses approximately in the range 
$M\in(10^{-23},10^{-19})\,M_\odot$ (i.e., $M\in(10^{10},10^{14})\,{\rm g}$), which will be our primary range of interest.\footnote{
Very recently, Ref.~\cite{Domenech:2023mqk} presented a complementary study related to the SGWB produced by even smaller PBHs (with mass 
$M_\text{\tiny PBH}\approx 10^{-28}\,M_\odot$) 
dominating the universe and responsible for reheating, and that could also potentially lead to DM relics and to a SGWB signal at 3G interferometers.}

The paper is structured as follows. In Sec.~\ref{sec:setup} we discuss the basic formulas we use to derive the constraints, 
based on threshold statistics describing compact object formation, second-order gravitational-wave~(GW) emission, and PBH evaporation.
We also present different scenarios that circumvent current cosmological and astrophysical constraints on Hawking emission, and discuss the important limiting effect of astrophysical foregrounds expected in the frequency range of ground-based GW detectors.
In Sec.~\ref{sec:results}, we summarize our results, showing the detectability prospects of ET in various scenarios. We conclude in Sec.~\ref{sec:conclusion}. 

\section{Setup}
\label{sec:setup}

\subsection{Power spectrum}
We consider formation scenarios where the spectrum of curvature perturbations is enhanced at small scales, triggering gravitational collapse.
In order to consider realistic 
narrow spectra as a benchmark, we will adopt the following functional form
\begin{equation}
 \mathcal{P}_\zeta(k) = 
 A_0 (k/k_0)^{n} \exp[2-2(k/k_0)^2],
 \label{eq:pzeta_PLexp}
\end{equation}
which is parametrized by the peak amplitude $A_0$, spectral growth index $n$, and the reference wavenumber $k_0$.
As a reference, we will consider a spectrum characterised by a growth $n=4$, yielding $\mathcal{P}_\zeta(k)\propto k^4$ for $k<k_0$, (see e.g.~\cite{Byrnes:2018txb,Franciolini:2022pav,Karam:2022nym,Ozsoy:2023ryl}), while it is Gaussian suppressed for $k>k_0$.
With this parametrisation, for $n=4$ the maximum is achieved at 
$\mathcal{P}_\zeta(k_0)=A_0$. 
Later on, we will also explore the possibility of a shallower growth at small $k$'s (assuming $n\simeq 1$), which enhances the low frequency tail of the associated SGWB.

\subsection{Computation of the abundance}

The first step is to consider the relationship between the horizon mass 
$M_H$ and the perturbation comoving wavenumber $k$~\cite{Franciolini:2022tfm},
\begin{equation}
    M_H \simeq  
 1.2 \times 10^{-25} M_\odot
 \left(\frac{g_*}{106.75}\right)^{-1/6}
 \left(\frac{k/\kappa}{10^{13} {\rm pc}^{-1}}\right)^{-2}.
 \label{M-k}
\end{equation}
The PBH mass resulting from the collapse is related to $M_H$ 
 by an order-unity factor controlled by the critical collapse parameters~\cite{Musco:2008hv}
 (see Ref.~\cite{Escriva:2021aeh} for a recent review).
Here, $g_*$ is the effective number of degrees of freedom of relativistic particles and was normalised to the value for the Standard Model at high energies. 
We keep track of the additional prefactor $\kappa \equiv k r_m $
that relates $k$ to the characteristic perturbation size at horizon crossing $r_m$~\cite{Germani:2018jgr,Musco:2018rwt,Escriva:2019phb,Young:2019osy}.
For definiteness, we will later on adopt the value $\kappa = 2.51$ that derives from assuming the spectrum \eqref{eq:pzeta_PLexp} \cite{Musco:2020jjb,Delos:2023fpm}.\footnote{Notice that in Ref.~\cite{Kapadia:2020pir,Kapadia:2020pnr,Romero-Rodriguez:2021aws} the factor $\kappa$ was fixed to unity. This systematically bias towards smaller peak frequencies the SGWB associated to a given PBH population. }

Adopting threshold statistics, we compute the mass fraction $\beta$ assuming Gaussian primordial curvature perturbations\footnote{
See, however, the recent Refs.~\cite{Ferrante:2022mui,Gow:2022jfb} for nonperturbative extensions of this computation if one assumes 
non-Gaussian primordial curvature perturbations. We will not explore this possibility in this paper. 
} and accounting for the nonlinear relationship between curvature and density perturbations~\cite{DeLuca:2019qsy,Young:2019yug,Germani:2019zez}.
One obtains
\begin{align}
\beta & 
= 
\mathcal{K}
\int_{\delta_l^{\rm min}}^{\delta_l^{\rm max}}
\d \delta_l
\left(\delta_l - \frac{3}{8}\delta_l^2 - \delta_c\right)^{\gamma}
P_\text{\tiny G}(\delta_l),
\label{eq:GaussianTerm1}
\\
P_\text{\tiny G}(\delta_l) & = \frac{1}{\sqrt{2\pi}\sigma(r_m)}e^{-\delta_l^2/2\sigma^2(r_m)},
\label{eq:GaussianTerm}
\end{align}
where $\delta_l$ is the linear (i.e. Gaussian) component of the density contrast, and the integration boundaries are dictated by having over-threshold perturbations and Type-I PBH collapse (see e.g.~\cite{Musco:2018rwt}).
We indicate with $\sigma(r_m)$ the variance of the linear density field computed at horizon crossing time and smoothed on a scale $r_m$ (see e.g. Ref.~\cite{Franciolini:2022tfm} for more details), while $\delta_c$ is the threshold for collapse.
We also introduced the parameters ${\cal K}$ and $\gamma$ to include the effect of critical collapse.
Overall, one finds the peak of the PBH mass distribution to be 
$M_\text{\tiny PBH} \approx 0.6 M_H$ (see e.g. \cite{Franciolini:2022tfm}).

\subsection{SGWB induced by adiabatic perturbations}

Large adiabatic perturbations are responsible for the GW emission at second order in perturbation theory~\cite{Tomita:1975kj,Matarrese:1993zf,Acquaviva:2002ud,Mollerach:2003nq,Ananda:2006af,Baumann:2007zm,Wang:2019kaf}.
The fundamental relation to consider is the one connecting the frequency $f_\text{\tiny GW}$ of the SGWB to the comoving wavenumber $k =  2 \pi f_\text{\tiny GW}$, which is 
\begin{align}\label{GW_peak_frequency}
f_\text{\tiny GW}
\simeq 15 \, {\rm kHz}
\left( \frac{k}{10^{13} /{\rm pc}} \right),
\end{align}
or equivalently, using Eq.~\eqref{M-k},
\begin{equation}\label{GW_peak_frequency2}
    f_\text{\tiny GW} = 
    4.1 \times 10^4\,  {\rm Hz}
    \lp\frac{\kappa}{2.51}\rp 
    \lp \frac{g_*}{106.75} \rp^{-1/12} 
    \lp\frac{M_H}{10^{-25} M_\odot} \rp^{-1/2}.
\end{equation}
The current energy density of 
GWs is given by 
\begin{align}
 \Omega_{\text{\tiny GW},0} & = 0.39\, \Omega_{r,0}
 \left(\frac{g_*(T_H)}{106.75}\right)
 \left(\frac{g_{*,s}(T_H)}{106.75}\right)^{-\frac{4}{3}}
 \Omega_{\text{\tiny GW},H}
 \label{eq:cg}
 \end{align}
 as function of their frequency $f_\text{\tiny GW}$, with 
 \begin{align}
 & \Omega_{\text{\tiny GW},H} = \label{eq:Transfer}
 \left(\frac{k}{k_H}\right)^{-2b}
 \int_{0}^{\infty}\!\!\d v
 \int_{|1-v|}^{1+v}\!\!\d u
 \mathcal{T}(u,v)
 \mathcal{P}_{\zeta}(ku)\mathcal{P}_{\zeta}(kv)
\end{align}
(see e.g. Ref.~\cite{Domenech:2021ztg} for a recent review).
Here, $b\equiv (1-3w)/(1+3w)$ with $w$ being the universe's equation of state parameter (the pressure to energy density ratio) at the emission time, $\Omega_{r,0}$ is the density fraction of 
radiation, $g_*(T)$ and $g_{*,s}(T)$ are the temperature-dependent
effective number of degrees of freedom for energy density and
entropy density, respectively, and $\mathcal{T}(u,v)$ is the transfer function 
\cite{Espinosa:2018eve,Kohri:2018awv}.
We denote with the subscript ``$H$'' 
the time when
induced GWs of the given wavenumber $k$
fall sufficiently within the Hubble horizon to behave as a radiation fluid in an expanding universe.
The characteristics of the SGWB emitted from a narrow spectrum of curvature perturbations can be summerised as follows. The low frequency tail scales like $\Omega_\text{\tiny GW} \simeq f^3$, due to the causality limited efficiency of the super-horizon emission~\cite{Caprini:2009fx,Cai:2019cdl,Hook:2020phx,Brzeminski:2022haa,Loverde:2022wih}.
On the other hand, one obtains a high frequency tail that depends on the drop-off of the curvature spectrum.

\subsection{PBH evaporation and remnants}
 The Hawking temperature of a static BH of mass $M_\text{\tiny PBH}$ is given, in natural units, by~\cite{Hawking:1975vcx} 
\begin{equation}\label{PBH_temp_hawk}
T_\text{\tiny PBH}= \frac{1}{8 \pi G M_\text{\tiny PBH}} = 
53\,  {\rm TeV}
\lp \frac{M_\text{\tiny PBH}}{10^{-25} M_\odot} \rp^{-1},
\end{equation}
where $G$ is Newton's constant.
The mass evolution of an evaporating static PBH follows (e.g.~\cite{Hooper:2019gtx,Inomata:2020lmk})
$ {{\rm d} M_\text{\tiny PBH}} / {{\rm d} t} 
 = 
 - {{\cal C}}/{M_\text{\tiny PBH}^2} 
$
where 
\begin{align}
	{\cal C} = \frac{\pi \, \mathcal G \, g_{\text{H}*}(T_\text{\tiny PBH}) M_\text{Pl}^4}{480}\,,
\end{align}
$\mathcal G \simeq 3.8$ is the gray-body factor, and $T_\text{\tiny PBH}$ is the PBH temperature in Eq.~\eqref{PBH_temp_hawk}.
$g_{\text{H} *}(T_\text{\tiny PBH})$ counts the the spin-weighted degrees of freedom of the particles produced from the Hawking radiation with $T_\text{\tiny PBH}$, whose concrete value ranges between 
$g_{\text{H}*} (T_\PBH) \approx 108 $ for $T_\PBH \gg 100 \, \text{GeV}$ and $g_{\text{H}*} (T_\PBH) \approx 7 $ for $T_\PBH \ll 1\, \text{MeV}$.
This can also be written as 
\begin{align}
 \frac{{\rm d} M_\text{\tiny PBH}}{{\rm d} t} 
 &= -9.7 \times 10^{-26} \, \frac{M_\odot}{\text{s}}
 g_{\text{H}*}(T_\text{\tiny PBH})
 \left( \frac{M_\text{\tiny PBH}}{10^{-25 }  M_\odot} \right)^{-2}.
 \label{eq:m_pbh_evo}
\end{align}
Solving Eq.~(\ref{eq:m_pbh_evo}), we can derive the lifetime of a static evaporating PBH
\begin{align}
  t_\text{\tiny eva} & =
  \frac{(M_\text{\tiny PBH}^{i})^3}{3 {\cal C}} 
  \approx 3.4 \times 10^{-3}\, {\rm s} \lp \frac{g_{\text{H}*}}{100}\rp^{-1} \lp \frac{M_\text{\tiny PBH}^i}{10^{-25 }M_\odot }\rp ^3. 
  \label{eq:mass-time_relation}
\end{align}
To simplify the interpretation of the previous equation, we have reported it in terms of the initial PBH mass, 
$M_\text{\tiny PBH}^i$, and have neglected a putative remnant mass, $M_f$ (see below).

For comparison, in a radiation-dominated FLRW universe where the Friedmann equation dictates
\begin{equation}
 H^2 = \frac{4 \pi ^3 }{45} g_*( T) G   T^4,   
\end{equation}
the age of the universe at a temperature $T$ is
\begin{equation}
t_\text{\tiny age} = 0.15\,  {\rm s} \lp \frac{g_*}{10.75}\rp^{-1/2} \lp \frac{T}{4 {\rm MeV}}\rp ^{-2}. \label{tage}
\end{equation}

One possibility we are interested in here is that PBHs do not evaporate completely but rather that, due to (possibly unmodelled) quantum gravity or anyway beyond-GR effects, Hawking evaporation leads to the formation of a stable remnant, of mass $M_f$ typically close to the Planck 
scale~\cite{MacGibbon:1987my,Barrow:1992hq,Carr:1994ar,Dalianis:2019asr,Chen:2014jwq}.
Anyway, in order to be agnostic let us assume that $M_f$ is a free parameter, not necessarily of the order of $M_{\rm Pl}$.
In the case such a remnant is formed, the timescale during which the evaporation is active is modified with respect to Eq.~\eqref{eq:mass-time_relation} as 
\begin{equation}
    t_\text{\tiny eva}=\frac{(M_\PBH^i)^3}{3{\cal C}} \llp 1-\frac{M_f^3}{(M_\PBH^i)^3}\rrp.
\end{equation}
Owing to a correction proportional to the third power of the mass ratio, only in the (fine-tuned and scale-dependent) case in which the remnant mass scale is very close to the initial PBH mass a sizable change to $t_\text{\tiny eva}$ is expected. In all other cases (including the most interesting one when $M_f=M_{\rm Pl} \ll M_\PBH^i$), the estimate~\eqref{eq:mass-time_relation} provides the time for the formation of a remnant.

\subsection{How to evade the BBN and CMB bounds}\label{sec:evadeBounds}
Before leaving a microscopic relic, Hawking emission might still lead to several effects in the early universe which should be taken into account.
In particular, for $M_\PBH\lesssim 10^{15}\,{\rm g}$, 
PBH evaporation is mostly constrained by big bang nucleosynthesis~(BBN) and distortions of the
cosmic microwave background~(CMB) (see e.g. Ref.~\cite{Carr:2020gox} for a detailed discussion on these and other constraints).
Below we consider various scenarios in which such constraints can be evaded or strongly relaxed, thus motivating the agnostic approach adopted in Sec.~\ref{sec:results}, where we will also take into account the possibility that non-GW constraints based on Hawking evaporation do not hold.

\subsubsection{Hawking remnants}
In the standard scenarios, a sizable fraction of the initial PBH mass evaporates before leaving a stable remnant.
Thus, to recover the predictions of standard big-bang cosmology, we require Hawking evaporation to terminate before BBN.
In practice, we impose $t_\text{\tiny eva}\leq t_\text{\tiny age}^\text{\tiny BBN}$, using Eq.~\eqref{tage} and assuming $T_\text{\tiny BBN}={\cal O}({\rm few})\, {\rm MeV}$.
This yields~\cite{Carr:2020gox}
\begin{equation}
    M^i_\text{\tiny PBH}\lesssim 5\times 10^{-24}\,M_\odot\,. \label{MBBN}
\end{equation}
Note that, owing to the cubic dependence in Eq.~\eqref{eq:mass-time_relation}, the mass corresponding to $t_\text{\tiny eva}= t_\text{\tiny age}^\text{\tiny BBN}$ is only mildly sensitive to ${\cal O}(1)$ changes in $T_\text{\tiny BBN}$ and even to larger changes in the parameters such as $g_{\text{H}*}$, $g_*$, and ${\cal G}$.
As previously discussed, the bound~\eqref{MBBN} is valid as long as $M_f\ll M^i_\text{\tiny PBH}$.

\subsubsection{Quasi-extremal PBHs} 
Beyond the possibility that the Hawking evaporation leaves behind a stable Planck mass remnant explaining the DM, one may also consider scenarios in which Hawking evaporation is not efficient.

One such possibility involves quasi-extremal PBHs (see Refs.~\cite{deFreitasPacheco:2023hpb} and references therein).
Indeed, quasi-extremal (charged and/or spinning) BHs in GR have nearly vanishing temperature $T_\text{\tiny PBH}$, which translates into a longer evaporation timescale and, most importantly, into a reduced particle emission rate (the latter being proportional to $T_\text{\tiny PBH}^4$).
We can define an extremal parameter $\epsilon$ 
that controls the BH temperature $T_\PBH= 1/(8 \pi M_\text{\tiny PBH})[4 \epsilon/(1+\epsilon^2)]$ and its luminosity $L\propto T^4_\PBH$.
For $\epsilon=1$ one recovers the Schwarzschild case while $\epsilon=0$ is the extremal case.
As a consequence, the BBN bounds and the CMB bounds~\cite{Carr:2020gox} in the relevant mass range can be evaded when~\cite{deFreitasPacheco:2023hpb}
\begin{equation}
    \epsilon\lesssim 10^{-3}\,.
\end{equation}
In this case quasi-extremal PBHs
remain stable on a time-scale comparable to the age of the universe and therefore represent viable DM candidates.

\subsubsection{PBHs in large extra dimensions}
It is likely that, in gravity theories introducing a new fundamental microscopic scale, $\ell$, Hawking evaporation of BH with size comparable to $\ell$ can be drastically different.

In theories with extra dimensions of size $\ell$, the evaporation timescale of a $d$-dimensional BH with horizon radius $r_0$ reads~\cite{Emparan:2000rs}
\begin{equation}
     t_\text{\tiny eva}^{(d)} \sim \left(\frac{\ell}{r_0}\right)^{2(d-4)}t_\text{\tiny eva}\,.
\end{equation}
Thus, provided $d>4$ and $r_0\ll \ell$, the evaporation time can be parametrically longer than Eq.~\eqref{eq:mass-time_relation}.
The different relation between mass, radius, and temperature in this scenarios provides different constraints on the PBH abundance~\cite{Friedlander:2022ttk}.
For example, for $d=6$ and extra dimensions at the $10\,{\rm TeV}$ scale, PBHs with mass $M_\PBH>10^{11}\,{\rm g}$ are viable DM candidates~\cite{Friedlander:2022ttk}.

\subsubsection{Ultraviolet GR deviations and minimum mass BHs}

A gravity theory with ultraviolet corrections to GR can be schematically written as
\begin{equation}
    S=\int d^4x \sqrt{-g}(R+ \ell^2 {\cal R}^2+...)
\end{equation}
where the first term is the standard Einstein-Hilbert action, ${\cal R}^2$ schematically denotes quadratic curvature corrections (implicitly including extra fields coupled to gravity), and $\ell$ is a new fundamental length scale below which GR deviations become dominant. Indeed, in various theories of this kind BHs have \emph{minimum} size and mass of the order $L$~\cite{Kanti:1995vq,Torii:1996yi,Pani:2009wy,Corelli:2022phw,Corelli:2022pio}.
At the same time, the Hawking temperature and the graybody factor of the minimum mass solutions are finite~\cite{Torii:1996yi,Konoplya:2019hml,Corelli:2022pio}, so the timescale and the very fate of Hawking evaporation in theories with high-curvature corrections is mysterious~\cite{Corelli:2022phw,Corelli:2022pio}.
These studies have been carried out in detail for Einstein-scalar-Gauss-Bonnet gravity~\cite{Kanti:1995vq} (a theory belonging to the Horndeski class and inspired by string theory), but are expected to be a generic feature of a new fundamental length scale (see, e.g., \cite{Casadio:2014pia}).
Interestingly, in Einstein-scalar-Gauss-Bonnet gravity the minimum-mass BH solution co-exists in the phase space of the theory with a regular wormhole without exotic matter~\cite{Kanti:2011jz}. This intriguing feature might suggest the possibility of a transition from this critical BH solution to a regular horizonless remnant, which does not evaporate any further~\cite{Corelli:2022phw,Corelli:2022pio}.

\subsubsection{Horizonless Relics}

Finally, as anticipated by the wormhole example just reported, the BBN and CMB constraints would not apply in case of formation of horizonless relics, for which Hawking radiation is totally absent or strongly suppressed~\cite{Raidal:2018eoo}.
Horizonless relics, cosmological solitons, and more generically exotic compact objects~\cite{Giudice:2016zpa,Cardoso:2019rvt} can form from large primordial curvature fluctuations~\cite{Raidal:2018eoo} and might be stable on cosmological timescales (see e.g.~\cite{Olle:2020qqy}).
Probably the best-known examples are Q-balls~\cite{Kusenko:1997si}, boson stars~\cite{Schunck:2003kk,Liebling:2012fv}, and oscillons~\cite{Seidel:1991zh}, which can form from the collapse of massive bosonic fields. 
Based on the nonlinear simulations in asymptotically-flat spacetime~\cite{Okawa:2013jba}, the threshold amplitude for the formation of a soliton can be much smaller than that for PBH formation, so the abundance of solitons is expected to be much higher.

Again in the spirit of an agnostic search, we shall parametrise the detectability of such scenario as a function of the collapse threshold $\delta_c^\text{\tiny HR}$ of these exotic compact objects (e.g.~\cite{Hidalgo:2017dfp}).
It is, therefore, convenient to define its ratio 
$R_{\delta_c} = \delta_c^\text{\tiny HR}/\delta_c^\text{\tiny PBH}$
with the threshold for PBH collapse in a radiation-dominated universe.

\subsection{Effects of expected astrophysical foregrounds}

To understand the detectability  of the SGWB signatures discussed in this work, it is important to assess the expected astrophysical foreground (AF) signatures falling within the same frequency range.
Focusing on ground-based experiments, the known source of GWs that is expected to dominate is represented by mergers of astrophysical binary BHs (BBHs) and neutron stars (BNSs). 
In particular, an astrophysical 
SGWB derives from the superposition of individually unresolved
GW emission from compact binary systems, as well as the residual background from imperfect removal of resolved sources.
The remaining SGWB foreground of astrophysical nature 
then limits how well we can observe the potentially subdominant cosmological SGWB discussed here.

\begin{figure}[!t] 
    \centering
\includegraphics[width=1.\columnwidth]{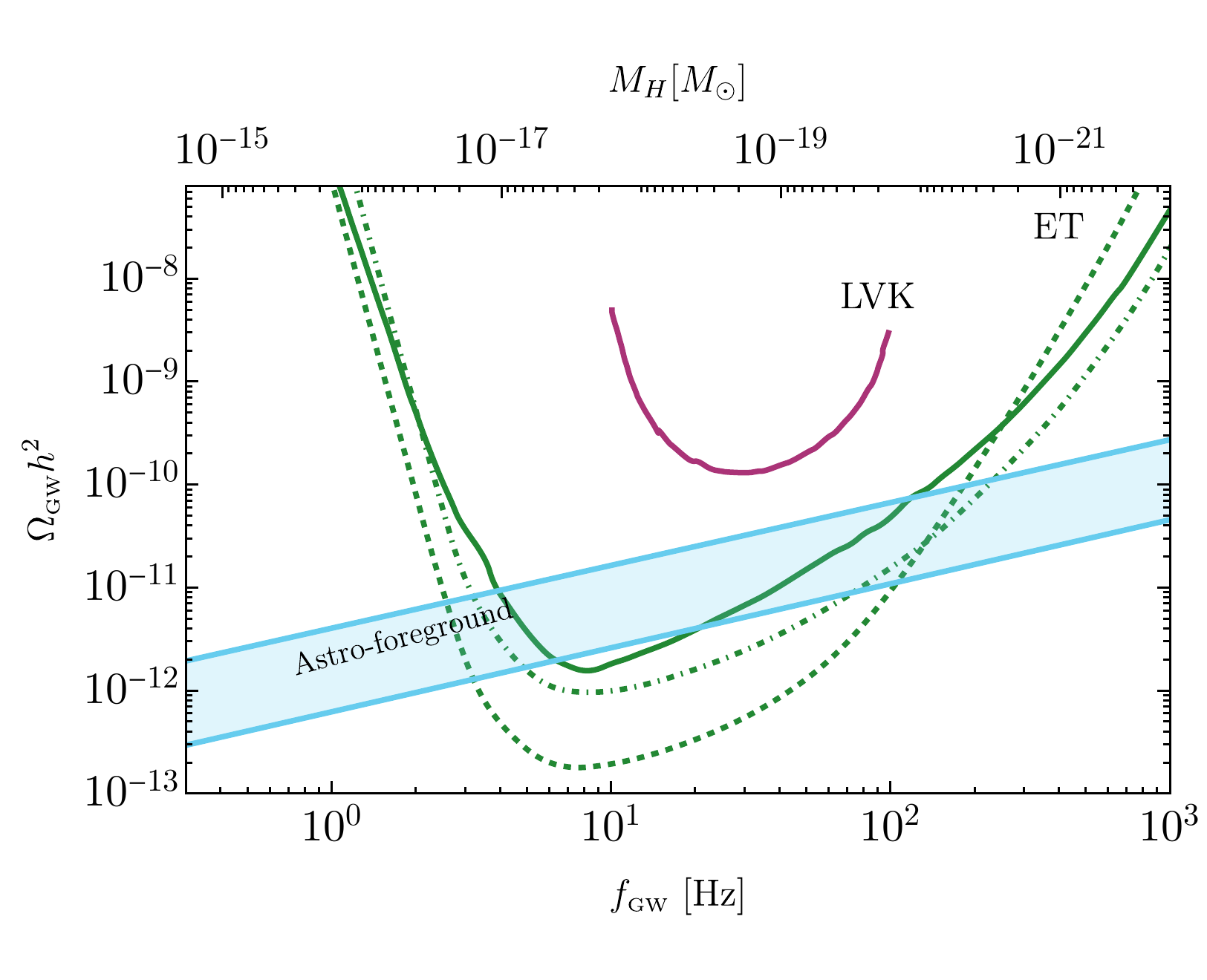}%
    \caption{GW interferometer sensitivities compared the astrophysical foreground from BBHs and BNSs.
 In particular, we show contribution from unresolved BNS as derived in Ref.~\cite{Zhou:2022nmt,Zhou:2022otw} (top line) 
    and \cite{Pan:2023naq} (bottom line), see more details in the main text.
    The green lines denote the integrated power-law sensitivity curves for $2\,{\rm yr}$ observation with ET in the following configurations (taken from Ref.~\cite{Branchesi:2023mws}): 
    i) single triangular detector ET-D (solid); 
    ii) single triangle, 15-km arms (dashed); 
    iii) two-detector network L-shaped 20-km arms (dot-dashed).
    The purple curve is the expected power-law integrated sensitivity curve obtained assuming 2-yr of observations with LIGO-Virgo-KAGRA (LVK) O5 at design sensitivity (see details on its derivation in Ref.~\cite{Bavera:2021wmw}). 
    }
\label{fig:astroforeground}
\end{figure}

\begin{figure*}[!t] 
    \centering
\includegraphics[width=1.6\columnwidth]{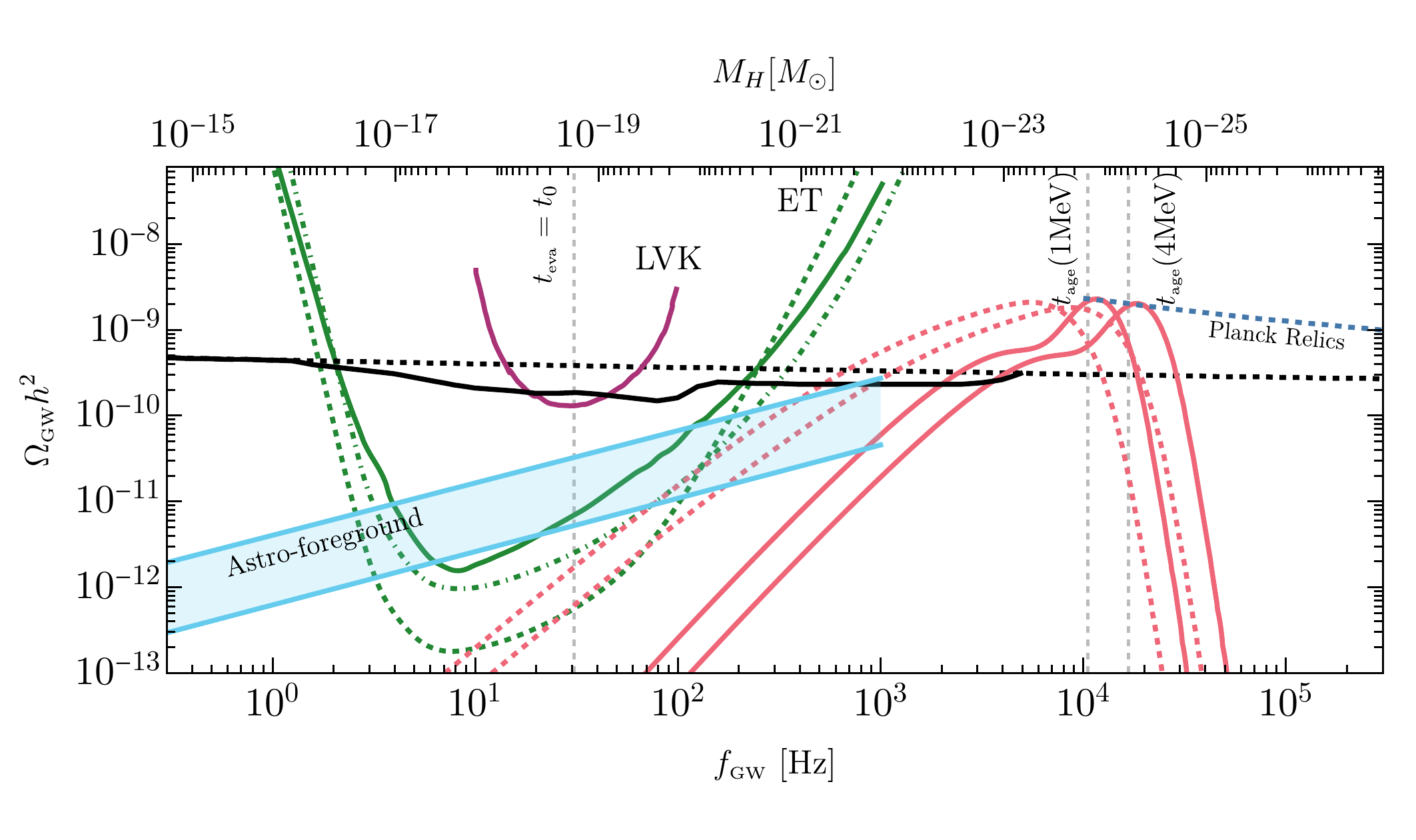}%
    \caption{ 
    The SGWB signal associated with PBH formation and comparison with ET sensitivity 
    (including the astrophysical background, see Fig.~\ref{fig:astroforeground}). 
    The nearly horizontal black dashed line indicates the  peak  spectral amplitude as a function of the varying peak frequency $f_\text{\tiny GW}$ associated with the formation of narrow PBH mass distribution with $f_\PBH=1$ neglecting bounds from Hawking evaporation (see Fig.~\ref{fig:SGWB_constraints_HR} below for details), while the solid black line denotes the corresponding  peak amplitude subject to the non-GW constraints from Hawking emission (see right panel of Fig.~\ref{fig:SGWB_constraints_3} below for details). 
    The red solid spectra correspond to the SGWB induced by the formation of PBHs evaporating before the onset of BBN and that could constitute the totality of the DM in the form of microscopic relics. Solid (dashed) red curves assume a steep (shallow) spectral growth with $n = 4$ ($n = 1$). 
    Finally, we indicate with ``Planck Relics'', the peak amplitude of SGWB for light PBHs that would evaporate leaving a Planck mass remnant. 
    }
\label{fig:SGWB_constraints}
\end{figure*}

The noise contribution from imperfect removal of resolved sources has been investigated in Refs.~\cite{Regimbau:2016ike,Sachdev:2020bkk,Martinovic:2020hru,Lewicki:2021kmu,Zhong:2022ylh,Zhou:2022nmt,Zhou:2022otw,Pan:2023naq,Smith:2017vfk,Sharma:2020btq}.
In Ref.~\cite{Pan:2023naq} in particular, at odds with previous claims based on different cleaning techniques, it was shown that adopting a method based on subtracting the approximate signal strain and removing the average residual power can reduce the noise from foreground BBHs and BNSs below the detector sensitivity limit. A sizable contribution still remains, however, from unresolved BNS mergers, which is still subject to uncertainties deriving from our limited knowledge of their intrinsic population based on currently available LVK data.
It should be noticed that this is, therefore, mostly insensitive to further improvements on
the techniques employed to subtract resolved signals (even though attempts were made in Refs.~\cite{Drasco:2002yd,Smith:2017vfk,Biscoveanu:2020gds}).

In Fig.~\ref{fig:astroforeground} we show the residual SGWB coming from {unresolved} BNSs
(as derived in Ref.~\cite{Zhou:2022nmt,Zhou:2022otw} (top line) and and \cite{Pan:2023naq} (bottom line) assuming a 3G detector network)
 normalised with the 90\% C.L. provided bu the population analyis of the LVK GWTC-3 data release~\cite{KAGRA:2021duu}. 
The upper side of the band is derived assuming the source-frame 
distribution of masses of each BNS to be described as in Ref.~\cite{Farrow:2019xnc}, 
where the primary mass follows a double Gaussian distribution and the secondary mass is sampled uniformly.
The merger rate redshift distribution should necessarily be extrapolated from the low $z$ observations, corresponding to approximately 
 $320/ ({\rm Gpc}^{3}{\rm yr})$ for BNSs. 
 A convolution of the star formation rate
 with a standard time-delay distribution $p(t_d) \approx 1/t_d$ was used, enforcing different minimum time delays for BBHs and BNSs~\cite{Dominik:2013tma,Vangioni:2014axa,LIGOScientific:2016fpe,LIGOScientific:2017zlf}.

\section{Results}
\label{sec:results}

\subsection{PBH evaporation and remnants}

In the first scenario, PBHs are formed with mass $M^i_\text{\tiny PBH}$ at early times, Hawking radiate for a time $t_\text{\tiny eva}$, and survive as stable remnants with mass 
$M_f$ (typically $\ll M^i_\text{\tiny PBH}$).
Due to the mass loss experienced by each PBH, the mass fraction in the PBH sector is decreased by a factor
$\beta_i/\beta_f = M^i_\text{\tiny PBH} / M_f$, which follows
from the conservation of the number density of objects. 
Therefore, a larger initial mass fraction is necessary in order for them to explain the entirety of the DM. However, as a consequence of the exponential dependence of the initial mass fraction to the perturbation amplitude (see Eq.~\eqref{eq:GaussianTerm1}),
this effect results in a larger SGWB that scales only weakly with $M_f/M_\text{\tiny Pl}$, where $M_\text{\tiny Pl}$ is the Planck mass. 
The SGWB peak amplitude assuming 
$f_\text{\tiny PBH} = 1$ for objects that
evaporate and leave a stable remnant at the mass scale $M_f$ is described by the following analytical fit
\begin{align}\label{eq:fitgw}
 &   h^2\Omega_\text{\tiny GW}(f)
    \simeq 
    10^{-8.1}
    \lp \frac{M_f}{M_\text{\tiny Pl}} \rp^{- 0.11}
    \times 
\nonumber \\ 
\big [&1-0.25 \log_{10}(f/{\rm Hz}) 
+ 1.6 \times 10^{-2}\log_{10}^2(f/{\rm Hz})
\big ].
\end{align}
 This fit is derived by varying $k_0$ in Eq.~\eqref{eq:pzeta_PLexp}, that corresponds to considering different scenarios of PBH formation with narrow mass faction peaked at the associated mass scale \eqref{M-k}. It should not be interpreted as a SGWB spectrum, but rather the largest amplitude that can potentially be reached by different SGWB spectra peaked at different frequencies. The shape of the SGWB spectrum in a given scenario, i.e. for fixed $k_0$,  features a low frequency $f^3$ tail dictated by causality \cite{Caprini:2009fx,Cai:2019cdl,Hook:2020phx}, a characteristic peak reaching the amplitude dictated by \eqref{eq:fitgw}, and a sharp decay at high large frequencies (see e.g. red spectra in Fig.~\ref{fig:SGWB_constraints}).
This amplitude is indicated in Fig.~\ref{fig:SGWB_constraints} as a dotted blue line denoted ``Planck relics'', assuming $M_f = M_\text{\tiny Pl}$.
As the effect of standard Hawking evaporation forces the 
characteristic mass of these objects to be lighter than the bound in Eq.~\eqref{MBBN}, in order to explain the entirety of the DM their SGWB must peak at frequencies larger than ${\cal O}(10^4)\,{\rm Hz}$.

We notice that, due to the exponential dependence of the abundance on the power spectral amplitude, changing the final mass of the remnant only implies a small change in the SGWB peak amplitude. 
For this reason, even though the amplitude of the SGWB 
associated with the Planck relic scenario is larger than the one expected for non-evaporating objects at masses below $M_\PBH\lesssim 5\times  10^{-24} M_\odot$,
its tail is not sizable enough to be visible by ET, as shown in Fig.~\ref{fig:SGWB_constraints}.
Also, the astrophysical foreground is prominent in the portion at large frequencies of the observable window of 3G detectors. Therefore, 
searching of signatures of even lighter compact objects would require the development of ultra-high frequency GW experiments, see e.g. Refs.~\cite{Aggarwal:2020olq,Franciolini:2022htd,Gehrman:2023esa}.

Next, we consider scenarios in which Hawking evaporation proceeds with a smaller efficiency.
The SGWB peak amplitude associated with scenarios that are able to explain $f_\text{\tiny PBH} = 1$ then follows the scaling 
\begin{align}\label{eq:fit_noevap}
    h^2\Omega_\text{\tiny GW}(f)
    \simeq & 
    4.5 \times 10^{-10} 
    \big [1-0.11 \log_{10}(f/{\rm Hz}) +
    \nonumber \\
    & 8.3 \times 10^{-3}\log_{10}^2(f/{\rm Hz})
    \big ],
\end{align}
valid within the reach of ground-based detector frequencies and shown as an almost horizontal, black dashed line in Fig.~\ref{fig:SGWB_constraints}. We remind the reader again that entering Eq.~\eqref{eq:fit_noevap} the peak frequency is associated with the characteristic PBH mass using Eqs.~\eqref{GW_peak_frequency}.
Such a SGWB background is only marginally visible within the 
O5 LVK design sensitivity\footnote{ Notice that we report the astrophysical contamination from unresolved BNS derived in Refs.~\cite{Zhou:2022otw,Pan:2023naq} assuming 3G detector sensitivities. One expects both a higher SGWB from unresolved sources and a far less efficient subtraction with LVK, resulting in a larger contamination.}
(with an observation time of 2 yr), 
while it would be fully accessible by ET.

\begin{figure*}[!t] 
    \centering
\includegraphics[width=\columnwidth]{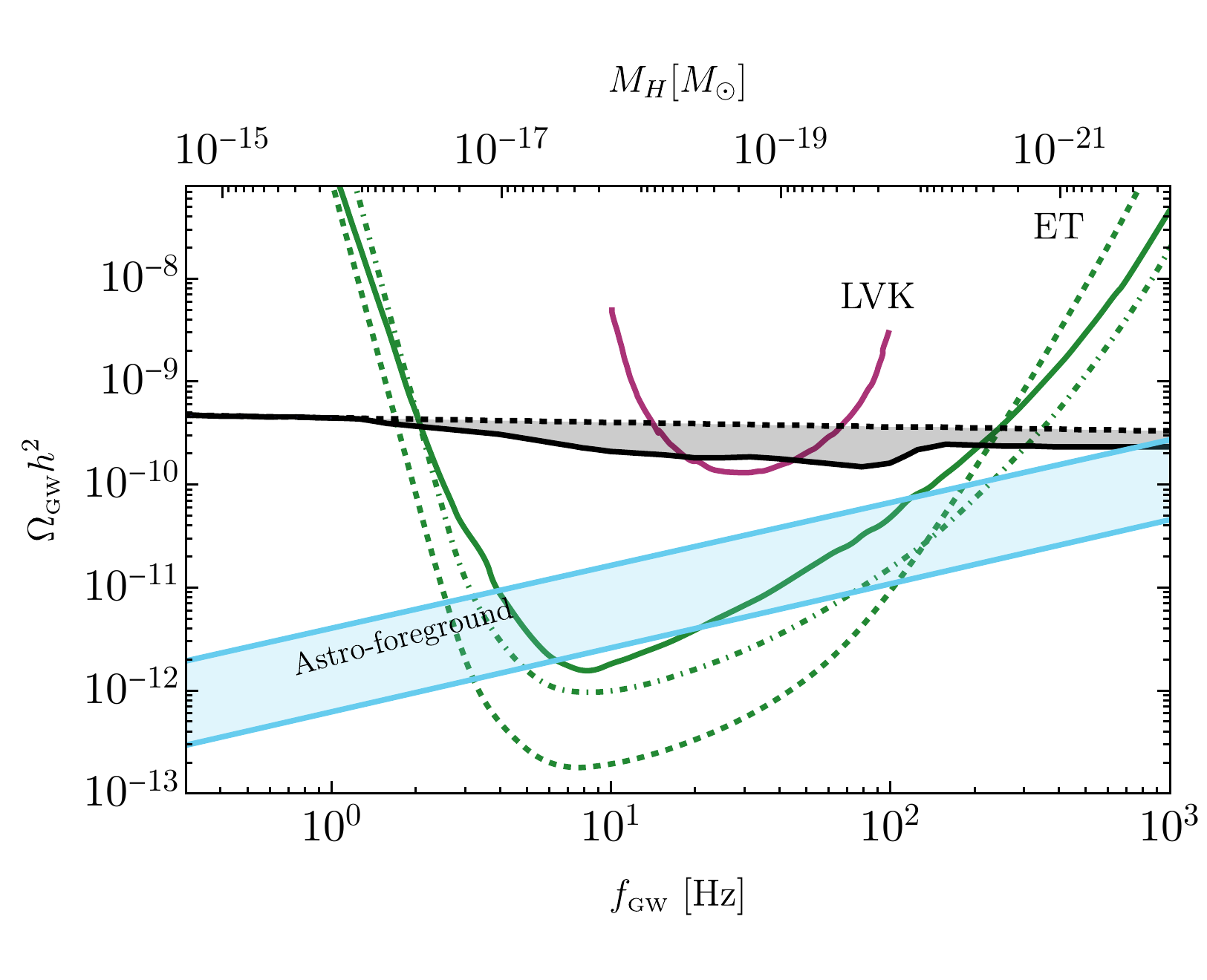}%
\includegraphics[width=\columnwidth]{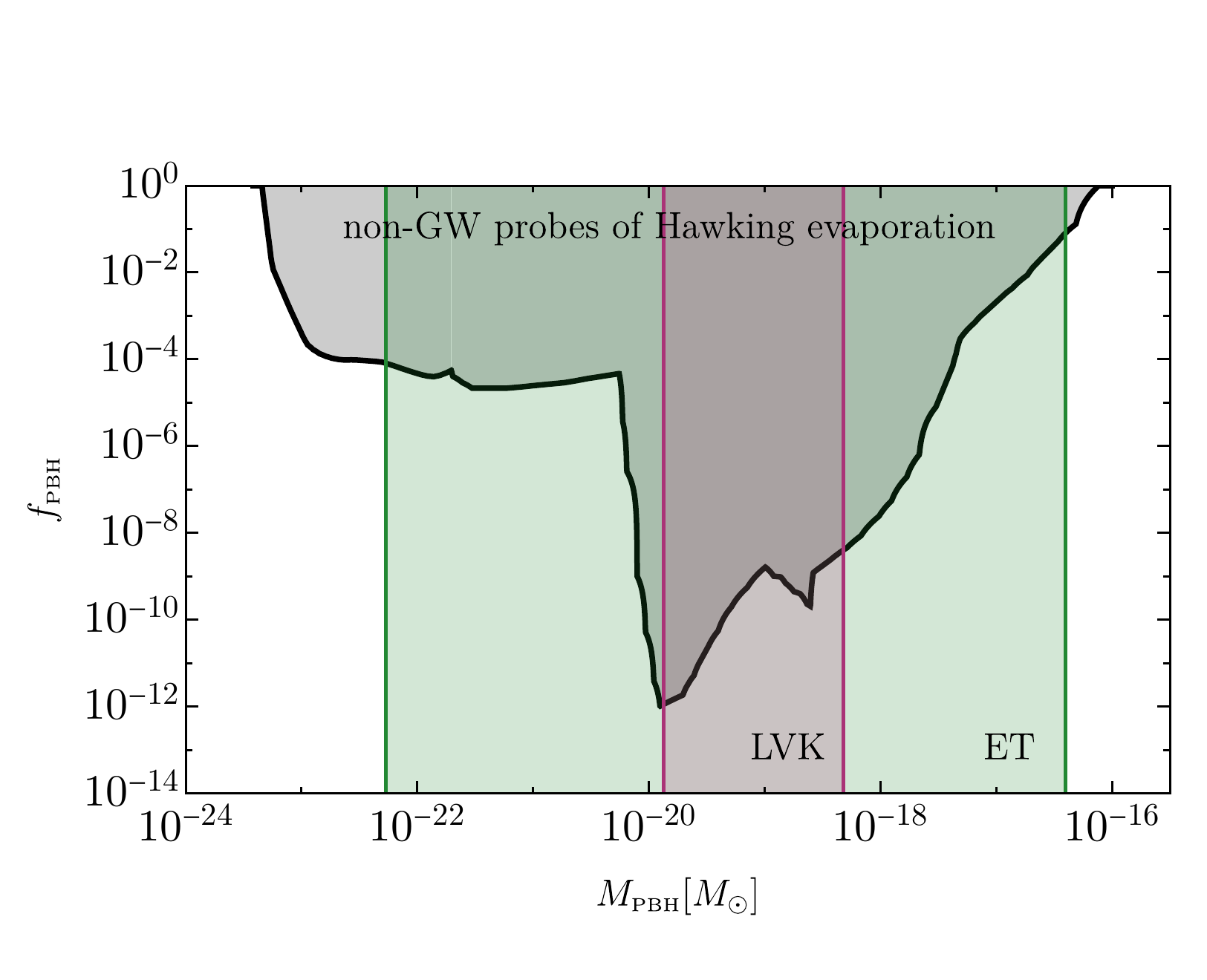}%
    \caption{ 
    \textbf{\textit{Left:}}
    Zoom-in of Fig.~\ref{fig:SGWB_constraints} (left panel) also detailing the astrophysical foregrounds (cyan band) dominated by unresolved BNS mergers.
    \textbf{\textit{Right:}}
    Range of mass where future ground-based detectors such as O5 LVK (purple) and ET (green) would be able to set constraints on the PBH abundance $f_\PBH$. In black, we shade constraints derived assuming Hawking evaporation is active and ruling out the effect of emission of SM particles on BBN, CMB, and late-time universe $\gamma$-ray observations~\cite{Carr:2020gox}.
    These plots assume a narrow PBH mass distribution (derived starting from the spectrum in Eq.~\eqref{eq:pzeta_PLexp}).
    }
\label{fig:SGWB_constraints_3}
\end{figure*}  

As these objects (discussed in Sec.~\ref{sec:evadeBounds}) are able to evade other constraints on their abundance coming from the effect of Hawking evaporation on BBN, CMB, and late-time electromagnetic observations, they represent a viable DM candidate, the formation of which could be probed by 3G detectors. 
In Fig.~\ref{fig:SGWB_constraints_3}, we show in detail the reach of ground based detectors such as LVK and ET. 
In the left panel, we report the SGWB peak amplitude that could be reached in these scenarios, where the dashed black line, corresponding to the fit \eqref{eq:fit_noevap}, saturates $f_\PBH = 1$. The region shaded in dark gray is where an observation would probe the existence of a population of compact objects whose Hawking emission 
is necessarily not efficient (or absent).
On the right panel, we show the range of masses where ET 
could constrain the fraction of DM in objects that do not evaporate, superimposed to non-GW constraints of the scenario that assumes standard Hawking evaporation.
The future bound coming from null observation of a SGWB of cosmological nature by ET would be incredibly stringent (i.e. $f_\PBH \ll 1$) but limited to the mass range 
\begin{equation}\label{eq.ETbound}
  M_\PBH\in 
    [5.3 \times 10^{-23} , 4.0 \times 10^{-17}]M_\odot.
\end{equation}
The bound reported in Eq.~\eqref{eq.ETbound} is shown in the right panel of Fig.~\ref{fig:SGWB_constraints_3}.
Finally, the right panel of Fig.~\ref{fig:SGWB_constraints_3} also shows the much larger range of frequencies (corresponding to a larger mass range) probed by the next generation of ground-based detectors compared to LVK.
 Given also that the LVK reach is limited by existing non-GW bounds, these results support the science case for ET.

\subsection{Horizonless relics}

ET would also be able to detect the SGWB associated with the formation of horizonless relics, avoiding bounds derived from Hawking evaporation effects shown in Fig.~\ref{fig:SGWB_constraints_3}.
The range of masses that could be probed, potentially 
being as extended as
$M_\text{\tiny HR} 
\in 
[5.3 \times 10^{-23}, 4.0 \times 10^{-17}] M_\odot $.
This range is however reduced in practice, if the threshold for formation of these objects was smaller than the one for PBHs.
In particular, we report results that assume different values for the ratio $R_{\delta_c} = \delta_c^\text{\tiny HR}/\delta_c^\text{\tiny PBH}$.

\begin{figure}[!t] 
    \centering
\includegraphics[width=\columnwidth]{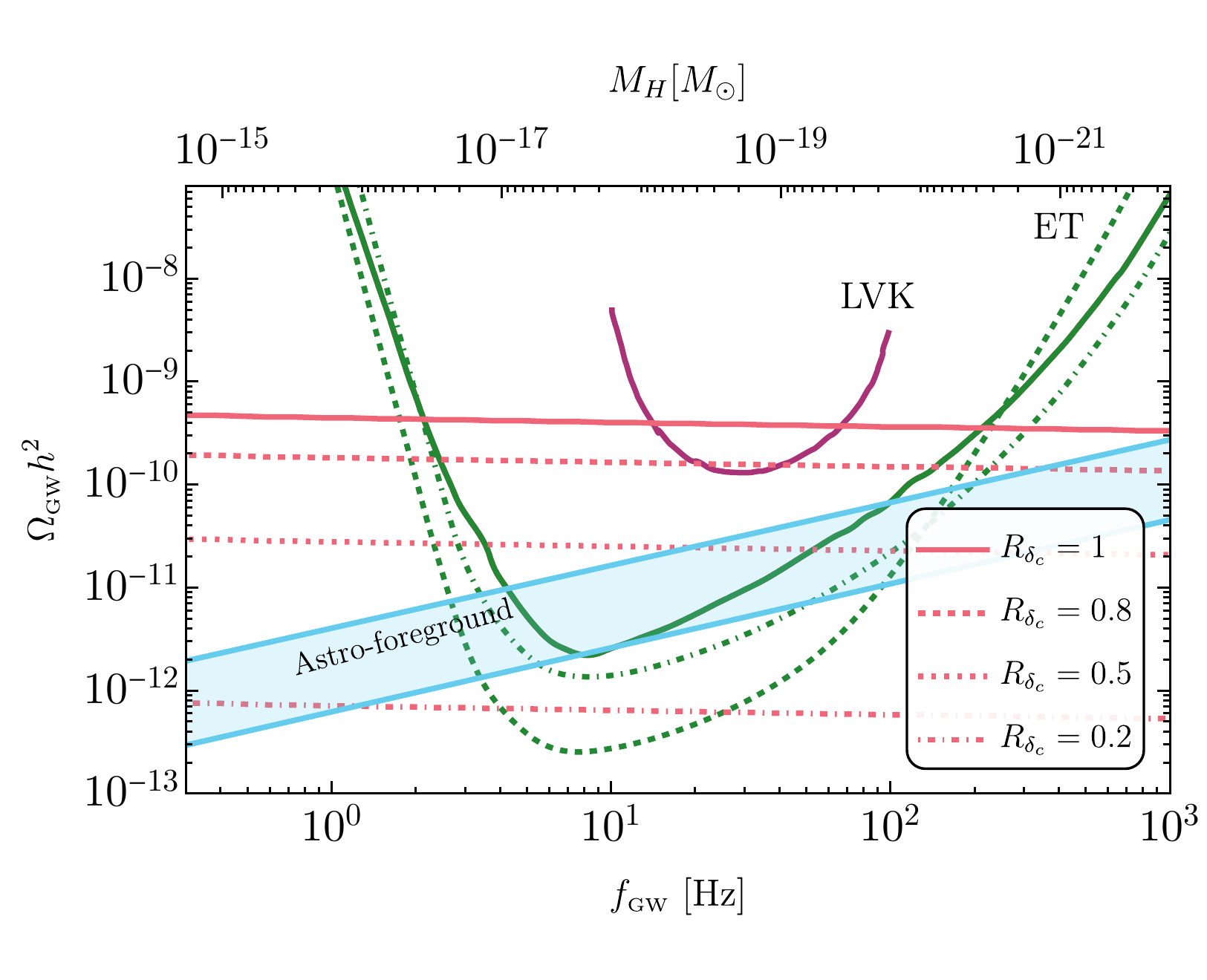}%
    \caption{ 
        Same as Fig.~\ref{fig:SGWB_constraints_3} (left panel), but considering 
    the case of horizonless relics that do not evaporate. 
    The red lines indicate the peak amplitude of the SGWB associated with the formation of non-evaporating compact objects.
    Each blue line corresponds to different assumptions on the threshold for formation of these objects, parametrised by a varying $R_{\delta_c}$.
    }
\label{fig:SGWB_constraints_HR}
\end{figure}

At first sight, we find that, fixing the amplitude of perturbations such that this putative population of objects made up the entirety of the DM, ET (in the 20-km L-shaped configuration~\cite{Branchesi:2023mws}) would only be able to observe signals if $R_{\delta_c} > 0.2$.
We summarise these prospects in Fig.~\ref{fig:SGWB_constraints_HR}.
In this case, however, an important limiting factor 
will be the presence of an astrophysical background, which further limits the reach of ET. 
Considering the median value for astrophysical background from unresolved BNS derived in \cite{Zhou:2022otw} (upper side of the cyan band), we see that for a narrow range of masses ET could potentially probe the formation mechanism of horizonless compact object whose threshold for formation is close to the one for PBHs, with $R_{\delta_c} \gtrsim 0.5$.

This estimate shows that the amplitude of perturbations associated with a detectable SGWB is comparable to that for PBH formation, despite the fact that horizonless remnants can in general form from much lower amplitudes, as in the case, e.g., of oscillons~\cite{Aurrekoetxea:2023jwd} (see also~\cite{Okawa:2013jba,
Lozanov:2019ylm}).
In practice, SGWB detectability at 3G interferometers always requires large curvature perturbations at small scales in the early universe.

\subsection{Other scenarios}

In this section, we discuss other scenarios of formation of exotic compact objects that could lead to the generation of a SGWB in the ET frequency range but would be distinguishable from the ones discussed in this paper.

So far, we assumed that the formation of compact objects takes place during a radiation dominated  era of the universe, and that such radiation domination lasts up the usual matter-radiation equality at redshift $z\approx 3400$. However, between the object formation  and the BBN era, different expansion histories are possible. 
In fact, an early matter-dominated era would result in a different relation between the BBN constraints and the PBH masses. 
As the temperature of the universe scales as $T \propto a^{-1}$, where $a$ is the scale factor of the universe, the number of $e$-efolds occurring between two different epochs can be related directly to the ratio of temperatures 
$N = \log (a_f/a_i) = \log (T_i/T_f) $ (neglecting the possible change of effective degrees of freedom).
An intermediate early matter-dominated phase has the effect of anticipating the onset of the BBN era, as $a\sim t^{1/2}$ during a radiation-dominated phase, while $a\sim t^{2/3}$ during an early matter-dominated era. 
For our purposes, it suffices to say that, in case such era is realised, the mass scale that is able to evade BBN bounds when Hawking evaporation is efficient
would shift to smaller PBH masses that evaporate faster.
Therefore, these scenarios would also not be visible with ground-based detectors. 

One may also consider the formation of PBHs themselves to take place during an early matter-dominated era of the universe. 
While the absence of radiation pressure drastically reduces the pressure for collapse, the effect of inhomogeneities and angular momentum of the collapsing overdensities force the spectrum of perturbations to be sizable to produce a significant population of compact objects~\cite{Harada:2016mhb,Harada:2017fjm,Kokubu:2018fxy,deJong:2021bbo,DeLuca:2021pls}. 
As shown in Refs.~\cite{DeLuca:2019llr,Dalianis:2020gup}, the evolution of density peaks 
leads to the generation of a time-dependent quadrupole that is responsible for the emission of a SGWB. 
Therefore, analogous signatures are expected in formation scenarios within the early matter domination, with an additional model dependence on the length of the matter domination phase, typically tracked using the reheating temperature $T_\text{\tiny rh}$.
The peak frequency of the SGWB is given approximately by the relation
\begin{align}\label{fMDpeak} 
 f_\text{\tiny GW} 
 \simeq  1.2 \times 10^4  \text{Hz}
\left(\frac{M_\text{\tiny PBH}}{10^{-25} M_\odot}  \right)^{-1/3} 
\left(\frac{T_\text{rh}}{10^{10}  \text{GeV}} \right)^{1/3}.
\end{align}
while the expression for the spectral energy density parameter  of the GWs at the peak frequency 
\begin{align}\label{Omegapeak}
\Omega_\text{\tiny GW} h^2
& \simeq 4.4\times 10^{-11}
\left(\frac{\sigma}{0.1}\right)
\nonumber \\
& \times 
\left(\frac{M_\text{\tiny PBH}}{10^{-25} M_\odot}\right)^{2/3} 
\left(\frac{T_\text{rh}}{10^{10} \text{GeV}}\right)^{4/3}.
\end{align}
The GW spectrum from this scenario would scales as $\Omega_\text{\tiny GW}(f)\propto f$ in the IR regime, 
while descending less steeply  $\Omega_\text{\tiny GW}(f) \propto f^{-1}$ in the UV, with additional oscillatory features around the peak~\cite{Dalianis:2020gup}. 
 This is a distinct feature that differs from the one produced during the radiation domination era~\cite{ Mollerach:2003nq, Ananda:2006af, Baumann:2007zm} or just before the transition to the radiation era~\cite{Inomata:2019ivs, Inomata:2019zqy}.  

Another scenario recently proposed in Ref.~\cite{Domenech:2023mqk} assume copious 
formation of PBHs that become the dominant component of the energy budget of the early universe (also dubbed PBH-domination era) and then decay to leave Planck mass remnant potentially contributing to the totality of the DM. 
As the SGWB in such scenario is induced by the isocurvature modes of Poisson nature in the PBH sector, its frequency is related to the PBH mass by \cite{Domenech:2020ssp,Domenech:2023mqk}
\begin{equation}
    f_\text{\tiny GW} 
    \simeq 
    0.53\,  {\rm Hz} 
    \lp \frac{M_\PBH^i}{10^{-25}M_\odot} \rp ^{-5/6}.
\end{equation}
This is much different from the one we adopt in Eq.~\eqref{GW_peak_frequency2} as it is controlled by the inverse of the PBH mean distance instead of the inverse horizon size at collapse. 
It is interesting, however, to notice that the SGWB is expected to feature a scaling $\Omega_\text{\tiny GW}(f) \propto f^{11/3}$ in the low-frequency tail, and a sharp drop-off beyond the peak, making it clearly distinguishable from the scenario discussed in this work.

\section{Discussion and Conclusion}
\label{sec:conclusion}
3G GW interferometers such as ET and Cosmic Explorer will have the unique opportunity to detect the SGWB associated with the formation of primordial microscopic compact objects which, in various scenarios, can survive until present times and explain the entirety of the DM.
In the relevant mass ranges these relics are microscopic and would hardly leave any direct signature, but the SGWB associated with their formation (or with their dynamics~\cite{Domenech:2023mqk}) would be a smoking gun for these DM candidates.
Although the details of various scenarios are either model dependent or somehow vague, a generic prediction is that non-GW bounds arising from Hawking radiation~\cite{Carr:2020gox} might be totally or partially evaded.

Interestingly, the mass range accessible by detecting the SGWB at 3G detectors largely overlaps with the range excluded by (non-GW) cosmological and astrophysical probes. Thanks to this coincidence of scales, 3G detectors could potentially test the absence or the suppression of Hawking evaporation for compact objects formed in the early Universe.
Conversely, even in the standard PBH scenario the absence of SGWB associated with PBH formation in this mass range would constrain the DM fraction in PBHs much better than current BBN and CMB constraints, even accounting for the contamination from an astrophysical foreground.

We also have highlighted that the exciting possibility of detecting the SGWB associated with microscopic DM relics may hinge on the ability to subtract the known foreground signal from ordinary binary coalescences of BHs and neutron stars, and might require a more accurate waveform modelling and new data-analysis techniques.

It is important to stress that the above opportunities are unique for 3G detectors since, even for the LVK network at design sensitivity, the signal is marginally detectable.

We also notice that there are two mass ranges which are still allowed by non-GW constraints, namely $M_\PBH\gtrsim10^{-16}\,M_\odot$ and $M_\PBH\lesssim 5\times 10^{-24}\,M_\odot$. In the former case PBHs would survive evaporation until present times (thus explaining all the DM) and their associated SGWB could be detectable by deci-Hz interferometers such as DECIGO (see, e.g.,~\cite{Inomata:2018epa,Arya:2023pod}). 
Also, a tail at large frequencies may be generated by a power spectrum that drops less steeply in the UV compared to Eq.~\eqref{eq:pzeta_PLexp}, which we assumed throughout this work, becoming potentially visible with ET.
In the latter range, evaporation ends before BBN and DM could be explained by Planck-mass remnants. The associated SGWB in this case peaks at $f_{\rm GW}>10^4\,{\rm Hz}$. Unfortunately, no planned high-frequency GW detector is expected to have the required sensitivity in that bandwidth~\cite{Aggarwal:2020olq}.

Interestingly, besides the connection with the DM problem, from a theoretical perspective, the fate of the Hawking evaporation and possible stable remnants are also relevant for the information loss paradox~\cite{Hawking:1975vcx,Mathur:2009hf,Polchinski:2016hrw,Chen:2014jwq}.

We have largely focused on the standard formation scenario from large adiabatic curvature perturbations, but other cases are possible, including SGWB from number density fluctuations of tiny PBH whose Hawking emission reheats the universe~\cite{Domenech:2023mqk}, or formation in an early matter dominated era of the universe~\cite{Dalianis:2020gup,Dalianis:2019asr}.
Likewise, horizonless compact objects require specific matter content so studying in detail their formation requires considering the collapse of extra matter fields and the SGWB associated with their isocurvature perturbations~\cite{Domenech:2021and}, as recently studied in~\cite{Lozanov:2023aez}. We plan to consider these extensions in the near future.

\begin{acknowledgments}
We are grateful to Emanuele Berti for interesting discussions and to Guillem Domenech for comments on the draft.
We acknowledge financial support provided under the European
Union's H2020 ERC, Starting Grant agreement no.~DarkGRA--757480 and support under the MIUR PRIN (Grant 2020KR4KN2 “String Theory as a bridge between Gauge Theories and Quantum Gravity”) and FARE (GW-NEXT, CUP: B84I20000100001, 2020KR4KN2) programmes.
G.F. acknowledges additional financial support provided by "Progetti per Avvio alla Ricerca - Tipo 2", protocol number
AR2221816C515921.
\end{acknowledgments}

\bibliography{main}

\end{document}